\documentclass[% 
reprint,
 amsmath,amssymb,
 aps,
]{revtex4-2}
\usepackage{graphicx}
\usepackage{dcolumn}
\usepackage{bm}
\usepackage{hyperref}
\usepackage{amsmath}
\usepackage{multirow}
\usepackage{tabularx}
\usepackage{booktabs}
\usepackage{caption}
\usepackage{float}
\usepackage{IEEEtrantools}
\usepackage{subcaption}

\usepackage[labelformat=empty]{subcaption}
\usepackage[section]{placeins} 
\usepackage{caption} 
\hypersetup{hypertex=true,
            colorlinks=true,
            linkcolor=blue,
            anchorcolor=blue,
            citecolor=blue,
            urlcolor=blue}
\usepackage{subcaption}
\usepackage{graphicx}
\usepackage{microtype}  
\usepackage{ragged2e}   
\usepackage{booktabs}
\usepackage{siunitx}
\begin{document}

\preprint{APS/123-QED}

\title{Microscopic Statistical Calculation of Nuclear Level Density Based on Relativistic Density Functional Theory}% Force line breaks with \\
\author{Zi-Cheng Wang}
 \affiliation{College of Physics, Jilin University, Changchun 130012, China}
\author{Peng-Xiang Du}
 \affiliation{College of Physics, Jilin University, Changchun 130012, China}
\author{Jian Li}\email{jianli@jlu.edu.cn}
\affiliation{College of Physics, Jilin University, Changchun 130012, China}
\author{ T. M. Shneidman}
\affiliation{Joint Institute for Nuclear Research, Dubna 141980, Russia}
\author{Shan-Gui Zhou}
\affiliation{Institute of Theoretical Physics, Chinese Academy of Sciences, Beijing 100190, China}
\affiliation{School of Physical Sciences, University of Chinese Academy of Sciences, Beijing 100049, China}
\affiliation{School of Nuclear Science and Technology, University of Chinese Academy of Sciences, Beijing 100049, China}


%\affiliation{
%\affiliation{
%\affiliation{%


\date{\today}% It is always \today, today,

\begin{abstract}
A microscopic statistical model based on the relativistic density functional theory (RDFT) is developed to calculate the nuclear level density (NLD). The approach employs self-consistent single-particle levels obtained from RDFT as input, incorporates pairing correlations within a finite-temperature Bardeen-Cooper-Schrieffer (BCS) theory, and accounts for rotational and vibrational collective enhancement effects. The spin cut-off parameter is calculated from the single-particle levels, thereby naturally retaining the shell effects and the structural characteristics of different nuclei. Using the shape-coexisting nucleus $^{98}\mathrm{Sr}$ as a representative example, the microscopic origin of the deformation effect on the NLD is investigated. In addition, the calculated NLDs are systematically compared with those from various phenomenological and microscopic models, as well as with available experimental data. The results indicate that although certain discrepancies exist among different models, they exhibit consistent overall evolutionary trends. Meanwhile, the RDFT-based microscopic statistical approach is capable of providing a reasonable description of the experimental NLDs as well as the $s$- and $p$-wave neutron resonance spacings.

%\begin{description}
%\end{description}
\end{abstract}

\maketitle

%\tableofcontents

\section{\label{sec:1}introduction}
Nuclear level density is a fundamental physical quantity that characterizes the statistical properties of nuclear excited states and plays a crucial role in the calculation of reaction cross sections \cite{RMP69244,PRC044602,NDS20093107,NST344}. It is also important for practical applications in reactor physics and nuclear medicine \cite{EPJWC2010212001,AIPCP20151653020076,EPJWC202226105008}. Therefore, obtaining accurate and reliable nuclear level density information is vital for improving our understanding of nuclear structure properties, advancing nuclear reaction theory, and supporting related scientific and technological applications.

However, the extraction of the nuclear level density from experimental data remains challenging. The most reliable experimental information is obtained from discrete levels at low excitation energies and neutron resonance spacings at the neutron separation energy. In addition, the Oslo method, developed in recent years, enables the simultaneous extraction of the $\gamma$-ray strength function and nuclear level density from $\gamma$-ray spectra measured in specific nuclear reactions \cite{NIMA498511,PRL232502l,EPJA5668,PRC014311}. Nevertheless, such experimental data are generally restricted to nuclei close to the valley of stability and to excitation energies below the neutron separation energy. For nuclei far from the $\beta$-stability line, at high excitation energies, or with special states such as fission saddle points, experimental access remains extremely limited. Consequently, the nuclear level density for these nuclei must be inferred from theoretical models.

In terms of theoretical models, early research was dominated by phenomenological approaches. Bethe first provided a simple analytical expression for the nuclear level density within the framework of the non-interacting Fermi-gas model \cite{PR50322,RMP69244}. Building on this foundation, a series of improved models were developed by incorporating shell corrections, pairing correlations, and collective effects, including the constant-temperature model (CTM) \cite{CJP14461496}, the back-shifted Fermi-gas model (BFM) \cite{NPA269298}, and the generalized superfluid model (GSM) \cite{PRC1993474}. Due to their simple form and computational efficiency, these models are widely used in nuclear reaction codes such as TALYS \cite{EPJA2023596} and EMPIRE \cite{NDS20071082655}. However, the parameters of these models rely on fitting existing experimental data, which limits their extrapolation capability and reliability in nuclear regions and energy ranges lacking experimental constraints \cite{EPJA169184}.

To address these challenges, a range of microscopic approaches have been developed. These include: the shell-model Monte Carlo method \cite{PRL42654268,PRC16781682,PRC034303,PRC852866,PRC836844}, the moments method based on random matrix theory and statistical spectroscopy \cite{PLB2018428,PRC024304}, the stochastic estimation method \cite{PLB20160370,PRC054306}, the Lanczos method employing realistic nuclear Hamiltonians \cite{PRC014315}, the projected shell model \cite{PRC034309}, the self-consistent mean-field approaches based on the extended Thomas–Fermi approximation with Skyrme forces \cite{PRC064302}, the exact pairing plus independent-particle model at finite temperature \cite{PRC054321,PRL022502,PRC054326,PLB634638}, the combinatorial method based on mean-field approximations \cite{EPJA169184,NPA6381,PRC064307,PRC064317,NPA913127156,NST34141,PLB138448,CPC2026505054107,arXiv20262605}, the microscopic statistical method based on mean-field approximations \cite{NPA640362,NPA60528,NPA19681101129,NPA69595,JNST48984,EPJA5097,PRC054315,PRC054606,NST34124}, the projected statistical method for restoring broken symmetries \cite{JMP1986276,PRL1993708}, and the quasiparticle random-phase approximation (QRPA) combined with boson-expansion technique \cite{PLB137989}. Of these, the microscopic statistical method offers a particularly effective balance between computational efficiency and accuracy, making it one of the most widely employed microscopic frameworks in current use. Within this framework, global microscopic calculations of the nuclear level densities were performed using the Hartree–Fock–BCS method in Ref. \cite{NPA69595}. In this approach, vibrational contributions to the level density are neglected. Level densities for several superheavy nuclei were also evaluated using a statistical method based on the two-center shell model in Ref. \cite{EPJA5097}. Furthermore, the level densities of Dy and Mo isotopes were calculated in Ref. \cite{PRC054315} using single-particle levels obtained from a Woods–Saxon potential.

The relativistic density functional theory has proven to be a powerful theory in nuclear physics and has been successfully applied to the analysis of both static and dynamic nuclear properties \cite{ZPA199133912313,PPNP199637193,PR2005409101,PPNP200657470,PRC2010821011301,PPNP201166519,PRC2012852024312,PLB20137264,NPP2015093101,Book2016,PS2016063008,PRL202212817,PRC20241106064319,PRC20251122L021303,PRC20251123}. It has several advantages due to the inclusion of the Lorentz symmetry \cite{PS20122012150}. In particular, it naturally includes the spin and pseudospin degrees of freedom simultaneously without requiring adjustable parameters, which can account for the large spin-orbit potential in nuclei and the origin of pseudospin symmetry \cite{PRL1997783,PRC1998582,PRC1999591,CPL2003203,PR20054144,PR20155701,PLB2024850138572}. Furthermore, time-odd fields are properly taken into account, which is essential for describing spectroscopic properties associated with nuclear rotations \cite{PRC2010823,FP201382095} and magnetic moments \cite{PRC19894031398,PLB19882143,PRC2006742024307,PRC2013886064307,SCSG200952101586,FP2018132109}. In recent years, RDFT is gradually being incorporated into microscopic statistical calculations of nuclear level density. For example, a microscopic method based on RDFT was developed in Ref. \cite{PRC054606}, in which the intrinsic level density is calculated from single-quasiparticle spectra obtained in finite-temperature self-consistent mean-field calculations, and the collective enhancement is determined from the eigenstates of a five-dimensional collective Hamiltonian. The method was further extended to odd-$A$ nuclear systems in Ref. \cite{NST34124} by introducing a core–quasiparticle coupling model. However, current microscopic statistical studies based on RDFT mainly focus on several typical nuclides and mostly consider only the excitation energy dependence of the level density. Systematic calculations and comprehensive evaluations across a broader range of nuclides that explicitly treat both spin and parity dependence are still lacking.

In this work, we develop a microscopic statistical approach based on RDFT to perform systematic calculations of nuclear level densities for even–even nuclei with available experimental $s$-wave neutron resonance spacings. This approach employs self-consistent
single-particle levels obtained from RDFT as input, incorporates pairing correlations within a finite-temperature BCS theory, and accounts for rotational and vibrational
collective enhancement effects. In addition,
particular attention is paid to the model-dependent behavior of the spin cut-off parameter and the influence of nuclear deformation on the level density.

The theoretical framework and methods of this study are introduced in Sec. \ref{sec:2}. Sec. \ref{sec:3} presents the results of spin cut-off parameters and level densities calculated using the statistical model based on RDFT, along with corresponding analysis and discussion. Finally, Sec. \ref{sec:4} provides the conclusions and future prospects.

\section{\label{sec:2}THEORETICAL FRAMEWORK}
Relativistic density functional theory starts from the Lagrangian density and constructs the energy density functional of nuclear systems within the mean-field approximation and the no-sea approximation. By minimizing the energy density functional, the Dirac equation for nucleons within the framework of relativistic mean-field theory can be obtained \cite{Book2016}:
\begin{equation}
\left[
\boldsymbol{\alpha}\cdot\boldsymbol{p}
+ \beta\big(m + S(\boldsymbol{r})\big)
+ V(\boldsymbol{r})
\right]\psi_k(\boldsymbol{r})
= \varepsilon_k \psi_k(\boldsymbol{r}) ,\label{eq:1}
\end{equation}
where $\boldsymbol{\alpha}$ and $\beta$ are the traditional $4\times4$ matrices of Dirac operators, $m$ is the nucleon mass, $S(\boldsymbol{r})$ and $V(\boldsymbol{r})$ are the local scalar and vector potentials, respectively, $\psi_k(\boldsymbol{r})$  is the corresponding single-particle wave function for a nucleon in the state $k$, and $\varepsilon_k$ is the corresponding single-particle energy.

To describe open-shell nuclei, pairing correlations are crucial. Starting from the Lagrangian density, a relativistic theory of pairing correlations in nuclei was developed in Ref. \cite{ZPA199133912313}. If we neglect the Fock terms as it is usually done in the relativistic density functional theory, the relativistic Hartree-Bogoliubov (RHB) equation for the nucleons reads
\begin{equation}
\begin{pmatrix}
h_D - \lambda_\tau & \Delta \\
-\Delta^{*} & -h_D^{*} + \lambda_\tau
\end{pmatrix}
\begin{pmatrix}
U_k \\ V_k
\end{pmatrix}
=
E_k
\begin{pmatrix}
U_k \\ V_k
\end{pmatrix},
\label{eq:2}
\end{equation}
where $E_k$ is the quasiparticle energy, $\lambda_\tau$ ($\tau = n, p$) are the chemical potentials for neutrons and protons, respectively, $h_D$ is the Dirac Hamiltonian in Eq. (\ref{eq:1}), $\Delta$ represents the pairing potential, and $U_k$ and $V_k$ are the quasiparticle wave functions.

In the present work, both spherical and axially symmetric deformed RDFT calculations are carried out using a harmonic-oscillator basis with 20 major shells, whose convergence has been verified. Pairing correlations are treated using a finite-range separable pairing interaction \cite{PLB200944,PRC064301}.

Based on the single-particle levels and pairing gaps calculated by RDFT, a microscopic statistical model within the framework of finite-temperature BCS theory is adopted to calculate the state density. In this approach, protons and neutrons are treated as two thermally equilibrated subsystems. Pairing correlations are taken into account through the BCS framework, and the state density is obtained using the saddle-point approximation \cite{NPA69595}. The state density expressed as a function of the excitation energy 
$U$ (see Eq. (\ref{eq:9})) is given by
\begin{equation}
\omega(U) = \frac{\exp \left[ S(U) \right]}{(2\pi)^{3/2} \sqrt{D}} , \label{eq:3}
\end{equation}
where $S(U)$ is the total entropy of the nuclear system, and $D$ is a determinant with its elements given in terms of the second derivatives of the grand partition functions of the nuclear system taken at the saddle point \cite{NPA197378}:
\begin{equation}
D = \left( \frac{1}{\beta^2} \frac{\partial^2 \ln Z_{p}}{\partial \lambda_p^2} \right) D_n + \left( \frac{1}{\beta^2} \frac{\partial^2 \ln Z_{n}}{\partial \lambda_n^2} \right) D_p , \label{eq:4}
\end{equation}
with
\begin{equation}
D_k = \begin{vmatrix}
\frac{1}{\beta^2} \frac{\partial^2 \ln Z_k}{\partial \lambda_k^2} & \frac{\partial}{\partial \beta} \left[ \frac{1}{\beta} \frac{\partial \ln Z_k}{\partial \lambda_k} \right] \\
\frac{\partial}{\partial \beta} \left[ \frac{1}{\beta} \frac{\partial \ln Z_k}{\partial \lambda_k} \right] & \frac{\partial^2 \ln Z_k}{\partial \beta^2}
\end{vmatrix} . \label{eq:5}
\end{equation}

The total entropy is the sum of neutron and proton contributions, $S(U) = S_{n}(U) + S_{p}(U)$, with
\begin{equation}
S_{k}(U) = 2\sum_{\nu} \left\{ \ln \left[ 1 + \exp(-\beta E_{k\nu}) \right] + \frac{\beta E_{k\nu}}{1 + \exp(\beta E_{k\nu})} \right\} , \label{eq:6}
\end{equation}
where $k = n,p$ represent neutrons and protons, respectively, and $\beta = 1/T$ is the inverse of the nuclear temperature. The quasiparticle energies are defined as $E_{k\nu} = \sqrt{(\varepsilon_{k\nu} - \lambda_k)^2 + \Delta_k^2}$, and $\varepsilon_{k\nu}$ are the single-particle energies obtained from RDFT. The Fermi energies $\lambda_k$ and pairing gaps $\Delta_k$ at different temperatures are determined self-consistently from the finite-temperature BCS equations:
\begin{equation}
N_{k} = \sum_{\nu} \left(1 - \frac{\varepsilon_{k\nu} - \lambda_k}{E_{k\nu}} \tanh \left[ \frac{1}{2} \beta E_{k\nu} \right]\right) , \label{eq:7}
\end{equation}
\begin{equation}
\frac{2}{G_k} = \sum_{\nu} \frac{\tanh \left[ \beta E_{k\nu} / 2 \right]}{E_{k\nu}} , \label{eq:8}
\end{equation}
where $G_k$ is the constant of the pairing interaction, whose value is determined from the quasiparticle energies at zero temperature.

The excitation energy $U$ is defined as the difference between the total energy of the system at temperature $T$ and its ground-state energy:
\begin{equation}
U = E(T) - E(0) , \label{eq:9}
\end{equation}
with the total energy given by
\begin{equation}
\begin{aligned}
E(T) = & \sum_{k=n,p} \bigg\{ \sum_{\nu} \varepsilon_{k\nu} \left( 1 - \frac{\varepsilon_{k\nu} - \lambda_k}{E_{k\nu}} \tanh \left[ \frac{1}{2} \beta E_{k\nu} \right] \right) \\
& - \frac{\Delta^2_k}{G_k} \bigg\} .
\end{aligned}
\label{eq:10}
\end{equation}

For nuclei in the spherical case, the spin-dependent level density is related to the spherical state density through
\begin{equation}
\rho_{\text{sph}}(U, J) = \frac{2J + 1}{2\sqrt{2\pi}\sigma^3} \exp\left[-\frac{J(J+1)}{2\sigma^2}\right]\omega_{\text{sph}}(U) . \label{eq:11}
\end{equation}

For deformed nuclei, within the hypothesis of axial symmetry, rotational degrees of freedom are taken into account, leading to a spin-dependent level density:
\begin{equation}
\begin{aligned}
\rho_{\text{def}}(U, J) &= \frac{1}{2} \sum_{K=-J}^{J} \frac{1}{\sqrt{2\pi\sigma^2}} \exp\left(-\frac{J(J+1)}{2\sigma_\perp^2}\right) \\
&\quad \times \exp\left[ -K^2\left(\frac{1}{\sigma^2} - \frac{1}{\sigma_\perp^2}\right)/2\right] \omega_{\text{def}}(U) .
\label{eq:12}
\end{aligned}
\end{equation}

The spin cut-off parameter $\sigma^2$ is obtained from the summation over the projections of the single-particle angular momentum on the symmetry axis \cite{NPA69595,PRC20074044308}:
\begin{equation}
\sigma^2(U) = \frac{1}{2} \sum_{k=n,p} \sum_{\nu} m_{k\nu}^2 \text{sech}^2 \left(\frac{E_{k\nu}}{2T}\right) , \label{eq:13}
\end{equation}
and the perpendicular spin cut-off parameter $\sigma_\perp^2$ affected by the pairing correlation is expressed as \cite{NPA69595}:
\begin{equation}
\sigma_\perp^2(U) = \sum_{k=n,p} \frac{\mathcal{I}_{\perp,k}}{\hbar^2} \left[1 - g\left(\frac{\delta\hbar\omega_0}{2\Delta_k}\right)\right] \times T \times \left(1 + \frac{1}{3}\delta\right) , \label{eq:14}
\end{equation}
with
\begin{equation}
g(x) = \frac{\ln(x+\sqrt{1+x^2})}{x\sqrt{1+x^2}} , \label{eq:15}
\end{equation}
where $\mathcal{I}_{\perp,k} = \frac{2}{5} m_{k}R^2$ is the moment of inertia with respect to the perpendicular axis,  $\delta = \beta_{2}/1.056$ is the deformation parameter and $\hbar \omega_{0} = 41A^{-1/3}\,\mathrm{MeV}$.

The contribution of vibrational collective modes is included through a vibrational enhancement factor:
\begin{equation}
K_{\mathrm{vib}}=\exp[\delta S-(\delta U/T)] , \label{eq:17}
\end{equation}
where $\delta S$ and $\delta U$ denote the changes in entropy and excitation energy, respectively, resulting from the vibrational modes. These changes are described by the Bose gas relationships. In this work, quadrupole, octupole, and hexadecapole vibrational modes are considered, and the related calculation process and expressions can be seen in Refs. \cite{NPA6381,PRC064307}. 

With increasing excitation energy, deformed nuclei tend to evolve toward a spherical shape. This shape transition significantly affects the nuclear level density. To account for this effect, a phenomenological damping function is introduced between the spherical and deformed expressions of the level density \cite{NPA60528,NPA69595,NPA6381}. The level density is then expressed as:
\begin{equation}
\begin{aligned}
\rho(U,J) = & f_{\mathrm{dam}}(U)K_{\mathrm{vib}}\rho_{\mathrm{def}}(U,J) \\
&+ [1-f_{\mathrm{dam}}(U)] K_{\mathrm{vib}}\rho_{\mathrm{sph}}(U,J) .
\end{aligned}
\label{eq:18}
\end{equation}
The damping function is defined as \cite{NPA60528}:
\begin{equation}
f_{\mathrm{dam}}(U)=\frac{1}{1+\exp[(U-{E}_{\mathrm{def}})/dU]} , \label{eq:19}
\end{equation}
where the deformation energy ${E}_{\mathrm{def}} = {E}_{\mathrm{sph}} - {E}_{\mathrm{eq}}$ is the energy difference between the spherical configuration and the equilibrium deformed state. The parameter $dU$ describes how fast the sphericity is recovered at energies above ${E}_{\mathrm{def}}$. In the present work, $dU = 6.27\,\mathrm{MeV}$ is determined by fitting experimental data at the neutron separation energy. This damping function is used to suppress the discontinuities that occur between the spherical and deformed level densities and to ensure a smooth transition from deformed to spherical shapes.

\section{\label{sec:3}Results and discussion}

\begin{figure}
\centering
\includegraphics[width=\columnwidth]{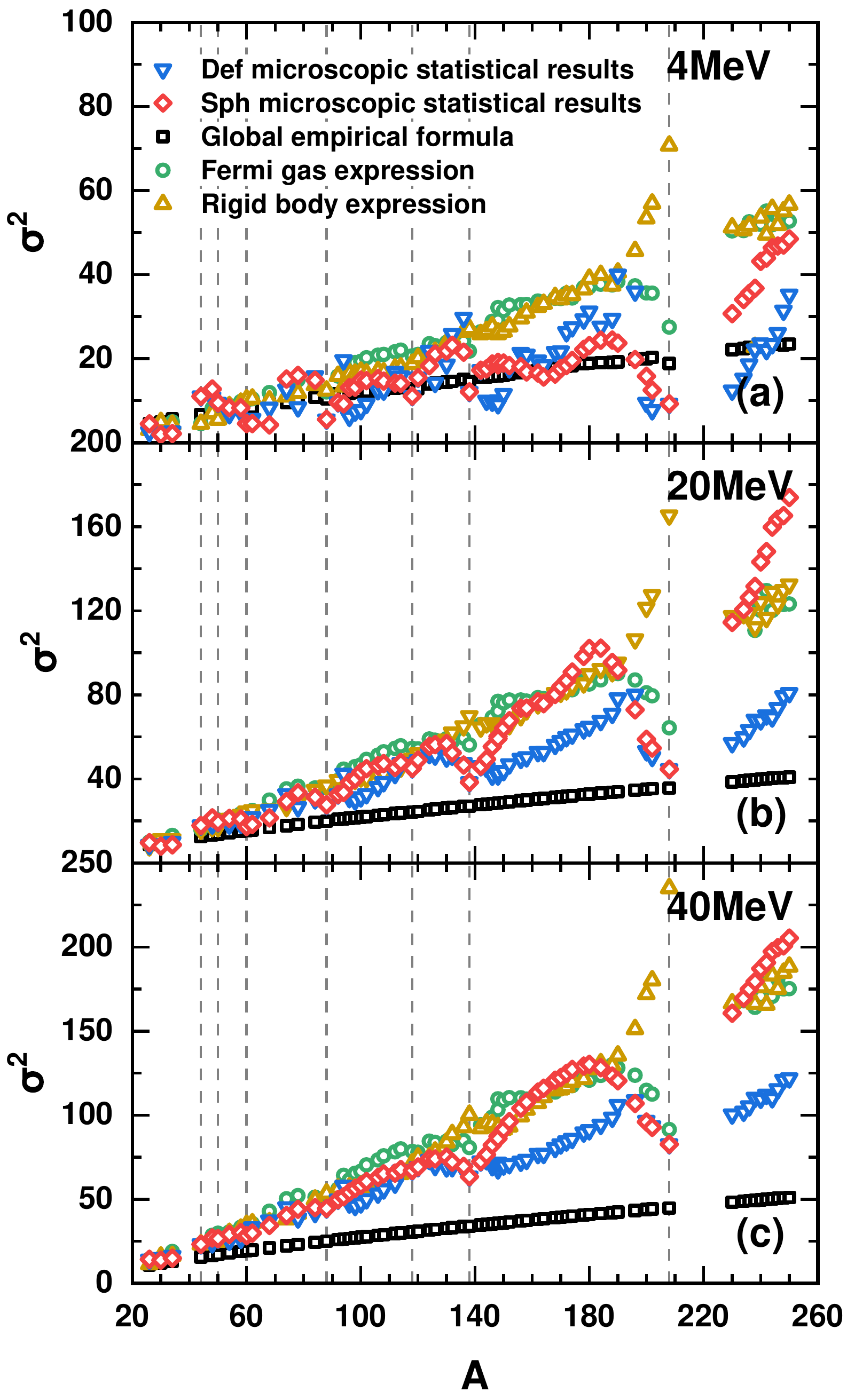}
\caption{\justifying Comparison of the spin cut-off parameters calculated using the microscopic statistical method (\ref{eq:13}), the global empirical formula (\ref{eq:20}), the Fermi-gas expression (\ref{eq:21}), and the rigid-body expression (\ref{eq:22}) at excitation energies of 4 MeV (a), 20 MeV (b), and 40 MeV (c). The dashed lines indicate the calculated results for $^{44}\mathrm{Ca}$, $^{50}\mathrm{Ti}$, $^{60}\mathrm{Ni}$, $^{88}\mathrm{Sr}$, $^{118}\mathrm{Sn}$, $^{138}\mathrm{Ba}$, and $^{208}\mathrm{Pb}$.}
\label{fig1}
\subcaptionbox{\label{fig:1(a)}}{}
   \subcaptionbox{\label{fig:1(b)}}{}
   \subcaptionbox{\label{fig:1(c)}}{}
\end{figure}

\subsection{Spin cut-off parameter}
The angular momentum distribution of the level density plays a crucial role in nuclear reaction calculations, and its width is characterized by the spin cut-off parameter $\sigma^2$. Therefore, an accurate description of the spin cut-off parameter is one of the key issues in statistical nuclear models.

To characterize the systematic behavior of the spin cut-off parameter, various empirical and semi-empirical formulas have been proposed \cite{PRC80054310,PRC102373}, including the global empirical formula obtained from fits to extensive experimental level schemes: 
\begin{equation}
\sigma^2 = 0.391A^{0.675}(U-0.5Pa')^{0.312} , \label{eq:20}
\end{equation}
the semi-empirical expression derived from the Fermi-gas model:
\begin{equation}
\sigma^2 = 0.1461\sqrt{a(U-\delta)}A^{2/3} , \label{eq:21}
\end{equation}
and the semi-empirical expression based on rigid-body rotational assumptions:
\begin{equation}
\sigma^2 = 0.0145\sqrt{(U-\delta)/a}A^{5/3} , \label{eq:22}
\end{equation}
where $U$ is the excitation energy, $Pa'$ is the deuteron pairing, $a$ is the level density parameter, and $\delta$ is the energy shift. The values of $Pa'$, $a$ and $\delta$ are taken from Ref. \cite{PRC80054310}.

Here, the spin cut-off parameters calculated using the microscopic statistical method [Eq. (\ref{eq:13})] with the DD-ME2 effective interaction \cite{CPC1856} are taken as an example and systematically compared with those obtained from Eqs. (\ref{eq:20})–(\ref{eq:22}). Figure \ref{fig1} shows the mass-number dependence of the spin cut-off parameters for 67 even–even nuclei at different excitation energies.

Figure \ref{fig:1(a)} shows that, at an excitation energy of 4 MeV, the spherical results are smoother than the deformed ones. This difference can be attributed to the splitting of single-particle levels induced by deformation, which leads to differences in the angular momentum distributions of the single-particle states near the Fermi surface that participate in the excitation. In addition, both microscopic statistical results exhibit local minima for nuclei near magic numbers. This behavior results from shell closures, which reduce the number of single-particle states available for excitation. In the same regions, the semi-empirical formula based on the Fermi-gas model also exhibits local minima, whereas the semi-empirical expression based on the rigid-body rotational assumption predicts local maxima. Overall, compared with the predictions of semi-empirical formulas, the microscopic statistical results show better agreement with the results of the global empirical formula.

As shown in Fig. \ref{fig:1(b)}, when the excitation energy increases to 20 MeV, the deformed results become smoother. This behavior arises from the increased number of single-particle states participating in the excitation at this energy, which reduces the differences in the angular momentum distributions of the single-particle states through statistical processing. In the medium- and heavy-mass region, the spin cut-off parameters for spherical case are generally larger than those for deformed case. This can be attributed to the higher degeneracy of single-particle levels in the spherical case, which provides more states available for excitation. At an excitation energy of 40 MeV [Fig. \ref{fig:1(c)}], the local minima in the light-mass region have basically disappeared. This is due to the gradual weakening of shell effects with increasing excitation energy. As can be seen from Figs. \ref{fig:1(b)} and \ref{fig:1(c)}, with increasing excitation energy, both the microscopic statistical results and the semi-empirical predictions exceed the results of the empirical fitting formula. This is because the empirical fitting formula is primarily based on experimental data at low excitation energies and therefore systematically underestimate the spin cut-off parameters at higher excitation energies.

\begin{table*}[t]
    \caption{Total energies and nuclear radii ($R$) of $^{98}\mathrm{Sr}$ at different deformations obtained from RDFT calculations. The proton and neutron Fermi energies ($\lambda_p$, $\lambda_n$) and pairing gaps ($\Delta_p$, $\Delta_n$) at $T = 0\,\mathrm{MeV}$, as well as the corresponding critical excitation energies ($U_{\mathrm{cr},p}$, $U_{\mathrm{cr},n}$), are obtained by solving the BCS equations.}

    \label{tab:1}
    \centering
    \begin{ruledtabular}
    \begin{tabular}{
        S[table-format=-1.2] 
        S[table-format=-4.4] 
        S[table-format=1.4] 
        S[table-format=-3.4] 
        S[table-format=-2.4]
        S[table-format=1.4]
        S[table-format=1.4]
        S[table-format=1.4]
        S[table-format=1.4]
        }
        \multicolumn{1}{c}{\textbf{$\beta_2$}} 
        & \multicolumn{1}{c}{\textbf{$\mathrm{Total}\ \mathrm{Energy}$}} 
        & \multicolumn{1}{c}{\textbf{$R$}} 
        & \multicolumn{1}{c}{\textbf{$\lambda_p$}} 
        & \multicolumn{1}{c}{\textbf{$\lambda_n$}} 
        & \multicolumn{1}{c}{\textbf{$\Delta_p$}} 
        & \multicolumn{1}{c}{\textbf{$\Delta_n$}} 
        & \multicolumn{1}{c}{\textbf{$U_{\mathrm{cr},p}$}} 
        & \multicolumn{1}{c}{\textbf{$U_{\mathrm{cr},n}$}} \\
        \multicolumn{1}{c}{\textbf{}} 
        & \multicolumn{1}{c}{\textbf{$\mathrm{(MeV)}$}} 
        & \multicolumn{1}{c}{\textbf{$\mathrm{(fm)}$}} 
        & \multicolumn{1}{c}{\textbf{$\mathrm{(MeV)}$}} 
        & \multicolumn{1}{c}{\textbf{$\mathrm{(MeV)}$}} 
        & \multicolumn{1}{c}{\textbf{$\mathrm{(MeV)}$}} 
        & \multicolumn{1}{c}{\textbf{$\mathrm{(MeV)}$}} 
        & \multicolumn{1}{c}{\textbf{$\mathrm{(MeV)}$}} 
        & \multicolumn{1}{c}{\textbf{$\mathrm{(MeV)}$}} \\
        \hline
        -0.60 & -803.3589 & 4.8193 & -14.1355 & -6.8440 & 1.0465 & 1.1190 & 4.7267 & 4.9059 \\
        -0.33 & -823.8499 & 4.5111 & -12.5097 & -4.9883 & 0.8529 & 0.0000 & 2.3594 & 0.0000 \\
        -0.20 & -822.2643 & 4.4643 & -13.1992 & -5.0060 & 1.2544 & 1.1513 & 5.8726 & 5.3133 \\
        0.20 & -821.0092 & 4.4584 & -12.9097 & -4.7728 & 1.3899 & 1.0034 & 7.4033 & 4.0714 \\
        0.46 & -823.7174 & 4.5372 & -13.3637 & -4.9360 & 0.0000 & 0.7809 & 0.0000 & 1.5589 \\
        0.60 & -820.9456 & 4.6443 & -13.8832 & -4.8901 & 0.8451 & 0.8375 & 2.6962 & 2.2570 \\
    \end{tabular}
    \end{ruledtabular}
\end{table*}

\begin{figure}
\centering
\includegraphics[width=\columnwidth]{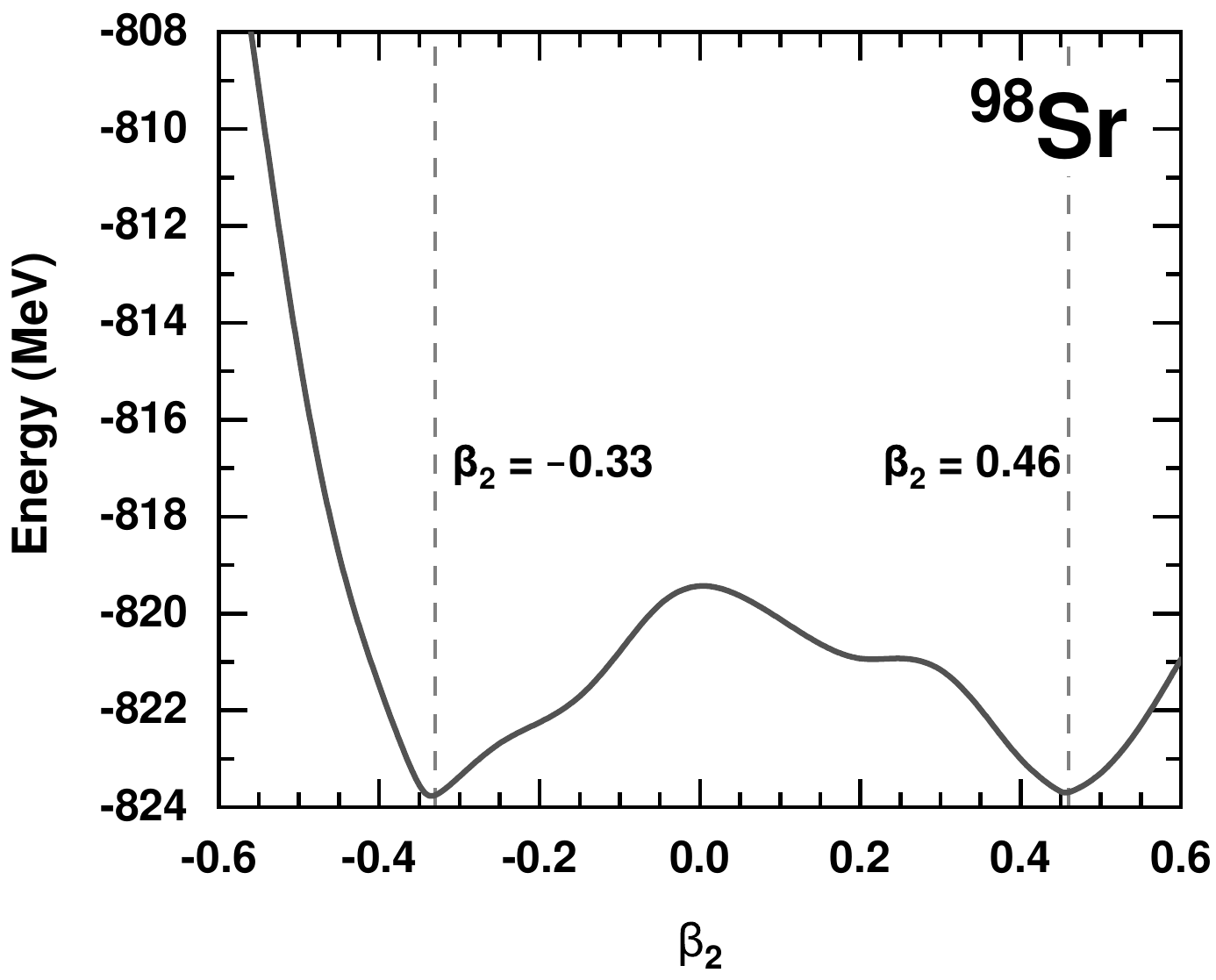}
\caption{\justifying Potential energy curve of $^{98}\mathrm{Sr}$. The dashed lines indicate the two minima located at $\beta_2 = -0.33$ and $\beta_2 = 0.46$.}
\label{fig2}
\end{figure}

\begin{figure}
\centering
\includegraphics[width=\columnwidth]{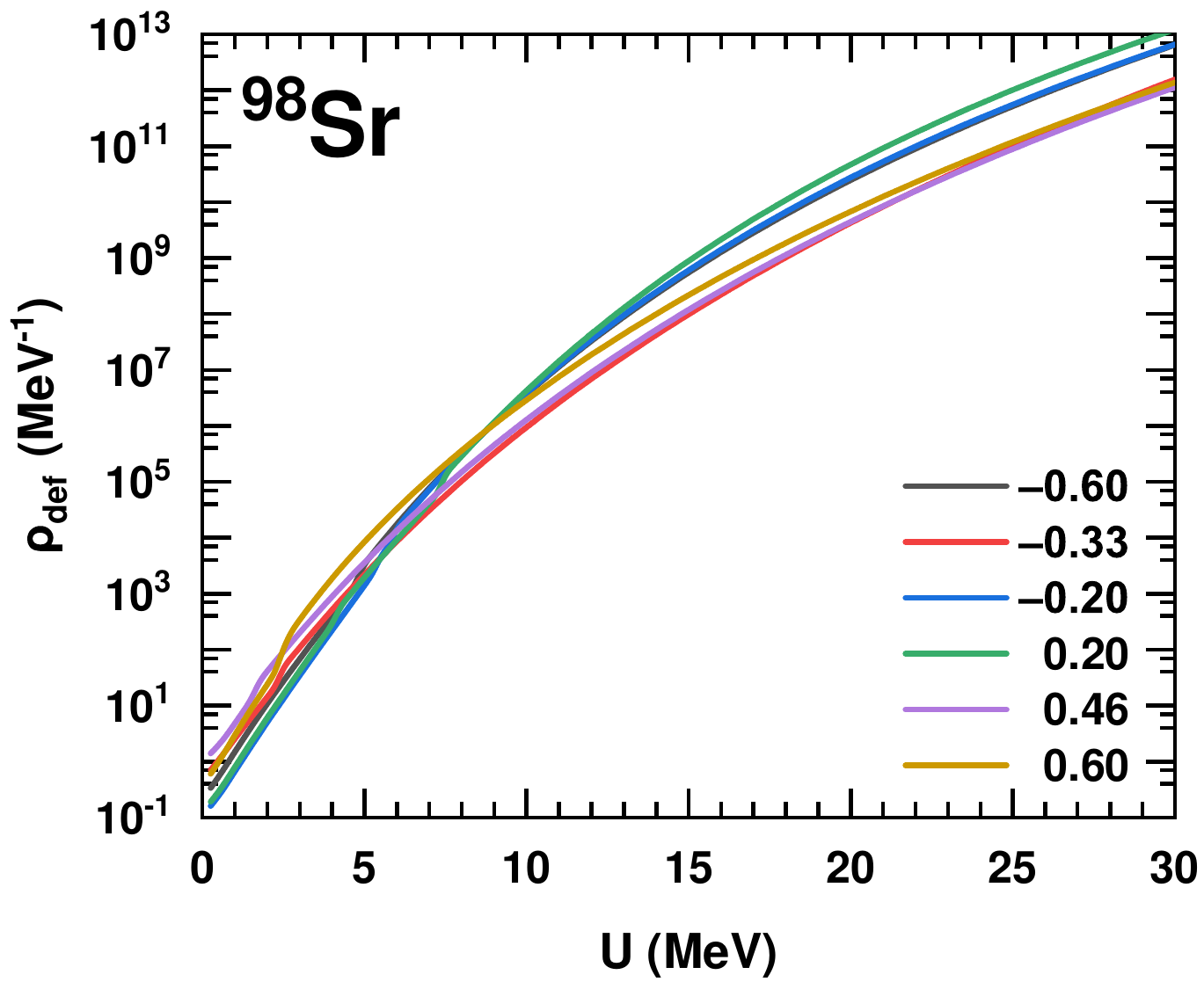}
\caption{\justifying Level densities of $^{98}\mathrm{Sr}$ for different deformations.}
\label{fig3}
\end{figure}

\begin{figure*}
\centering
\includegraphics[width=16cm]{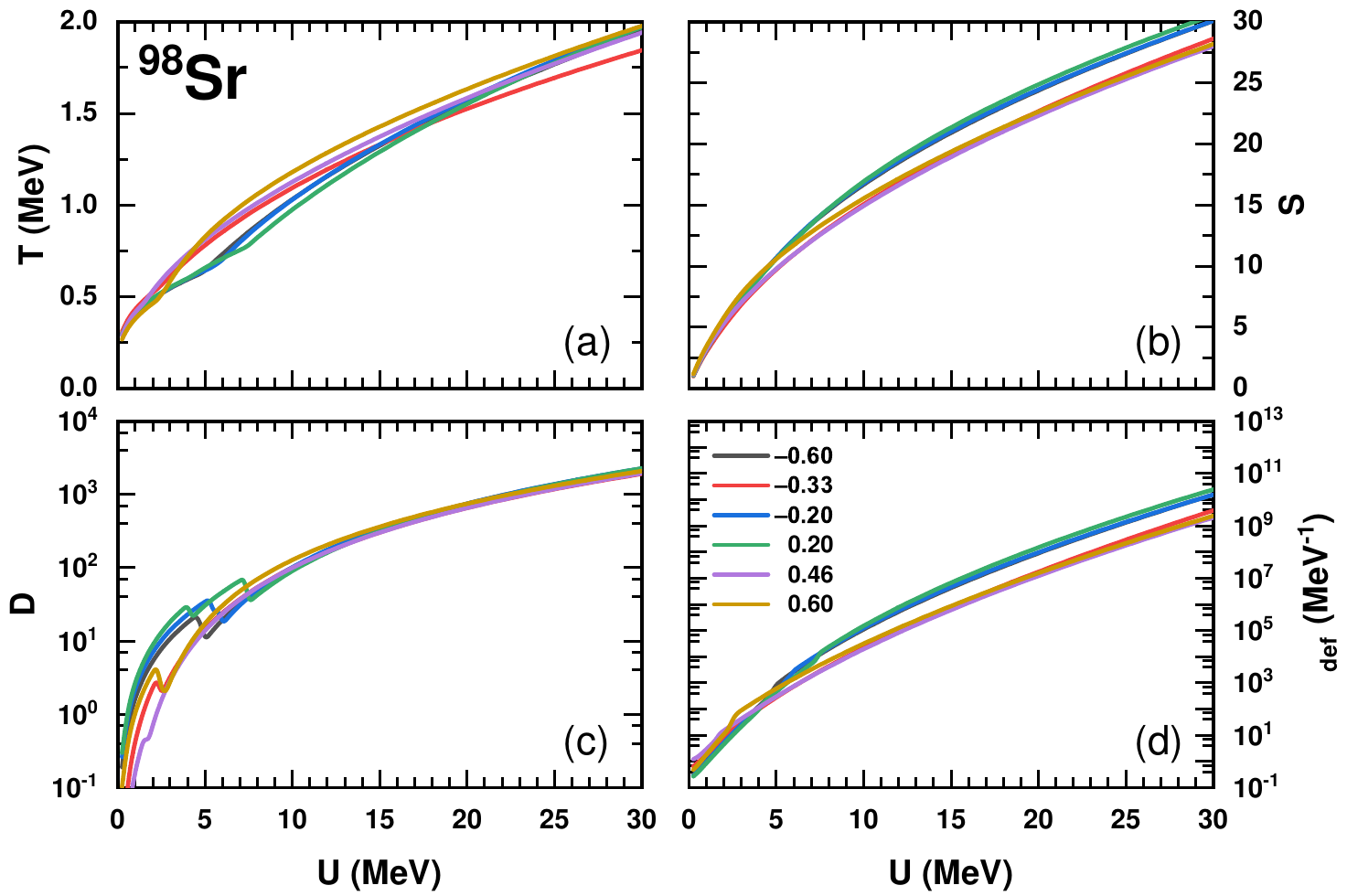}
\caption{\justifying Temperature (a), entropy (b), $D$ value (c), and state density (d) of $^{98}\mathrm{Sr}$ for different deformations as functions of excitation energy.}
\label{fig4}
\subcaptionbox{\label{fig:4(a)}}{}
   \subcaptionbox{\label{fig:4(b)}}{}
   \subcaptionbox{\label{fig:4(c)}}{}
   \subcaptionbox{\label{fig:4(d)}}{}
\end{figure*}

\begin{figure*}
\centering
\includegraphics[width=16cm]{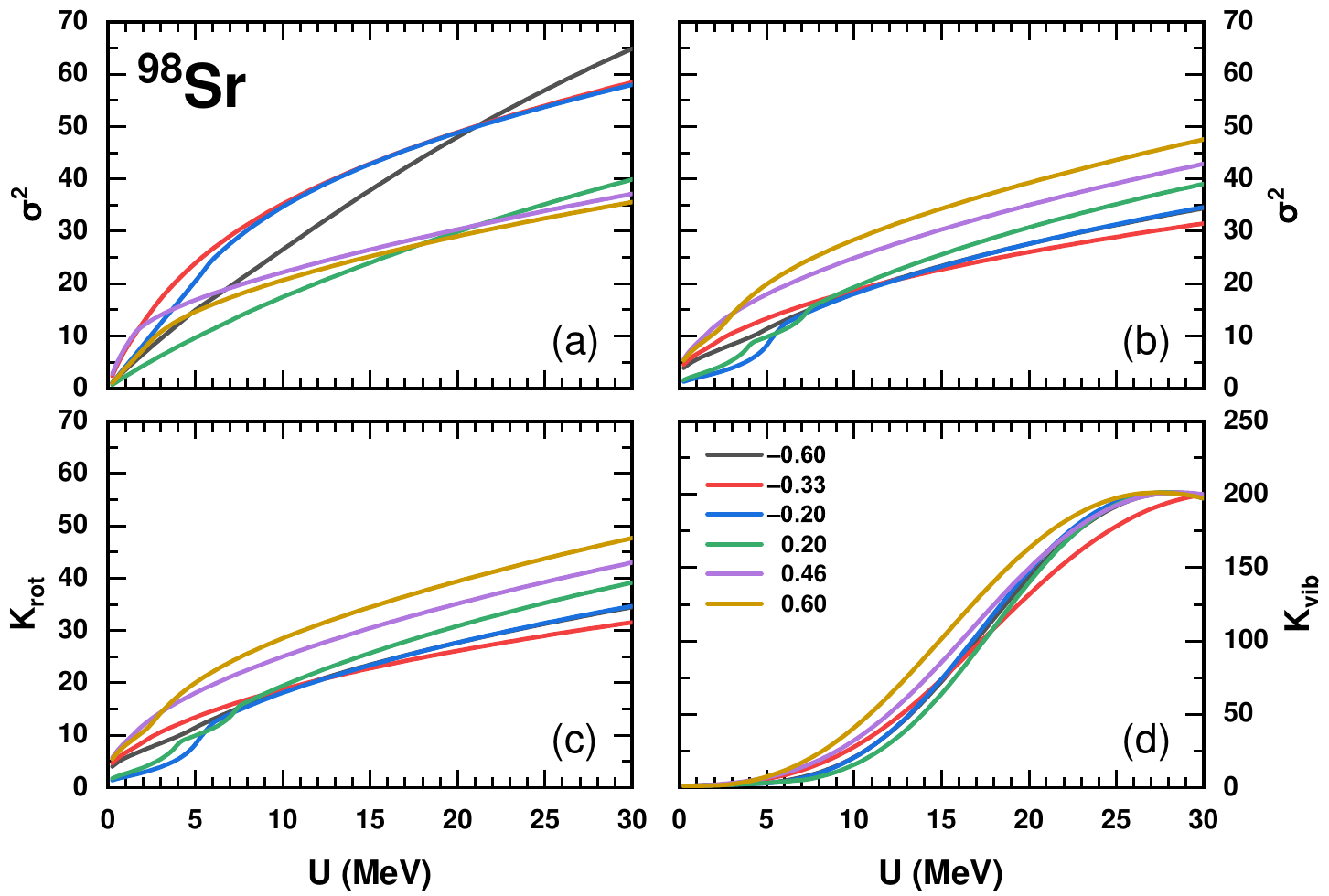}
\caption{\justifying Spin cut-off parameter(a), perpendicular spin cut-off parameter(b),
rotational enhancement factor(c), and vibrational enhancement factor(d) of $^{98}\mathrm{Sr}$ for different deformations as functions of excitation energy.}
\label{fig5}
\subcaptionbox{\label{fig:5(a)}}{}
   \subcaptionbox{\label{fig:5(b)}}{}
   \subcaptionbox{\label{fig:5(c)}}{}
   \subcaptionbox{\label{fig:5(d)}}{}
\end{figure*}

\subsection{Level densities of $^{98}\mathrm{Sr}$ for different deformations}

Figure \ref{fig2}  shows the potential energy curve of $^{98}\mathrm{Sr}$ calculated using the DD-ME2 effective interaction. Two pronounced minima are observed at $\beta_2=-0.33$ and $\beta_2=0.46$, with an energy difference of only about $0.13~\mathrm{MeV}$. This small energy difference indicates a typical shape-coexistence phenomenon in $^{98}\mathrm{Sr}$. Based on this result, we perform systematic calculations of the level densities of $^{98}\mathrm{Sr}$ at different deformations. The deformation dependence of the level density and its microscopic origin are then analyzed.

The level densities corresponding to different deformations are shown in Fig. \ref{fig3}, where significant differences among the various deformations are clearly observed. In this work, the level density is obtained by incorporating collective rotational and vibrational effects on top of intrinsic excitations. To better understand the deformation dependence, we separately analyze the state density and the collective enhancement effects. This analysis helps to clarify the physical origin of the differences in the level density associated with nuclear deformation.

As listed in Table \ref{tab:1}, the pairing gaps of $^{98}\mathrm{Sr}$ exhibit noticeable differences for different deformations at $T = 0\,\mathrm{MeV}$. Figure \ref{fig4} presents the excitation energy dependence of the temperature, entropy, $D$ value, and state density for different deformations. As shown in Fig. \ref{fig:4(a)}, when the excitation energy is below the critical energy of the pairing phase transition, the temperatures corresponding to different deformations are very similar. Once the excitation energy exceeds the critical value, the temperatures corresponding to different deformations begin to diverge. This behavior arises from the differences in the pairing gaps. At the same excitation energy, deformations with larger pairing gaps correspond to lower temperatures. This temperature difference gradually diminishes with increasing excitation energy. From Table \ref{tab:1}, it can be seen that at the two shape-coexisting deformations $\beta_2=-0.33$ and $\beta_2=0.46$, one of the neutron or proton pairing gaps is zero, while the non-zero pairing gaps are relatively similar. However, as the excitation energy increases, the temperature corresponding to $\beta_2=-0.33$ becomes significantly lower than those at other deformations. This behavior is due to the fact that, in the energy range from 7.5 to 15 MeV above the neutron Fermi surface,  the number of single-particle states at $\beta_2 = -0.33$ is significantly higher than those at other deformations. Consequently, with increasing excitation energy, more neutron single-particle states participate in the excitation at this deformation. As shown in Fig. \ref{fig:4(b)}, the entropies for different deformations are very similar at low excitation energies. According to  $T^{-1} = \partial S/\partial U$, the temperature difference essentially reflects the different growth rates of entropy with excitation energy. As the excitation energy increases, the temperature differences gradually vanish, and the rate of entropy growth under different deformations tends to become uniform. Figure \ref{fig:4(c)} shows that in the low excitation energy region, the $D$ values exhibit pronounced differences among different deformations. These differences arise from variations in the pairing gaps, and deformations with larger pairing gaps correspond to larger $D$ values. After the pairing correlations are broken, the $D$ values for all deformations gradually converge. As shown in  Fig. \ref{fig:4(d)}, since the state density increases exponentially with entropy, the differences in the state density at low excitation energies are mainly governed by the $D$ values, whereas at higher excitation energies they are dominated by the entropy differences. Indeed, the density of states is closely related to the distribution of single-particle levels near the Fermi surface—namely, the single-particle level density. In general, the larger the single-particle level density, the larger the state density. It should be noted that the single-particle level density near the Fermi surface is higher at $\beta_2 = -0.60$, $-0.20$, and $0.20$ than at $\beta_2 = -0.33$, $0.46$, and $0.60$. Consequently, once pairing correlations are broken, the state densities for the former set gradually exceed those of the latter. At low excitation energies, the larger state densities associated with $\beta_2 = -0.33$, $0.46$, and $0.60$ originate from their smaller pairing gaps compared to the other deformations.

Figure \ref{fig5} presents the excitation energy dependence of the spin cut-off parameter, the perpendicular spin cut-off parameter, the rotational enhancement factor, and the vibrational enhancement factor for different deformations of $^{98}\mathrm{Sr}$. As shown in Figs. \ref{fig:5(a)} and \ref{fig:5(b)}, at high excitation energies after the pairing correlations are broken, the spin cut-off parameter increases, whereas the perpendicular one decreases, as the deformation changes from prolate to oblate. It should be noted that for excitation energies above the critical energy but below approximately $20~\mathrm{MeV}$, although $\beta_2=-0.60$ is more oblate in shape than $\beta_2=-0.20$, its spin cut-off parameter is smaller. This behavior originates from the presence of single-particle states with larger magnetic quantum numbers near the proton Fermi surface at $\beta_2 = -0.20$. As the excitation energy further increases, more single-particle states become involved, and the spin cut-off parameters gradually return to the global trend associated with deformation. As shown in Table \ref{tab:1}, the total energy and nuclear radius at $\beta_2=-0.60$ are significantly larger than those of other deformations. After the pairing correlations are broken, the perpendicular spin cut-off parameter at $\beta_2=-0.60$ eventually exceeds that at $\beta_2=-0.33$ as the excitation energy increases (Fig. \ref{fig:5(b)}).

\begin{figure*}
\centering
\includegraphics[width=16cm]{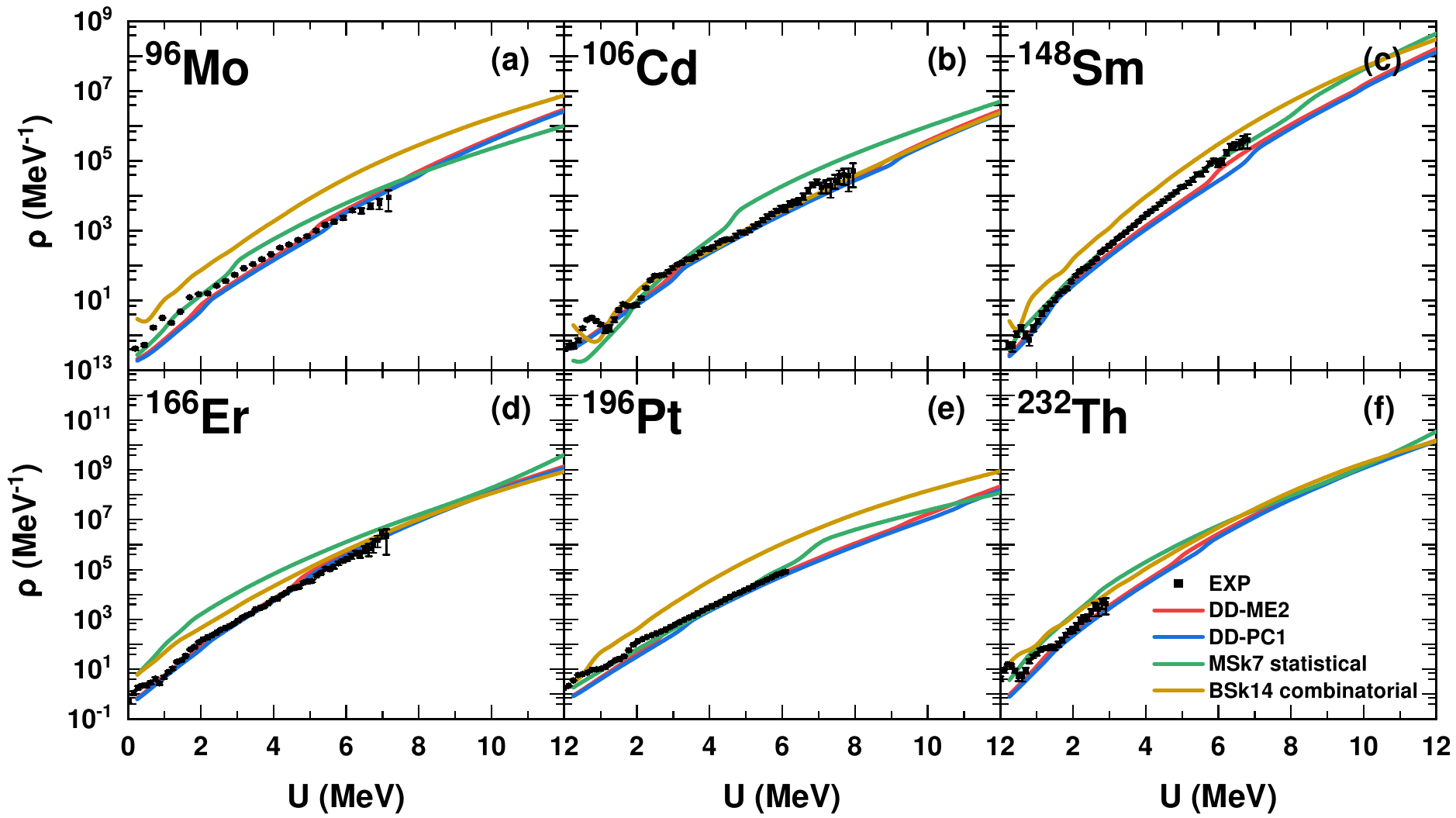}
\caption{\justifying Comparison of NLDs from the RDFT-based method with those from non-relativistic microscopic models and available experimental data. The RDFT results are calculated using the DD-ME2 and DD-PC1 effective interactions. The non-relativistic results from the Hartree-Fock-BCS plus statistical model with MSk7 and the Hartree-Fock-Bogoliubov plus combinatorial method with BSk14 are taken from the TALYS code \cite{EPJA2023596}. The experimental data were extracted using the Oslo method.}
\label{fig6}
\end{figure*}

\begin{figure*}
\centering
\includegraphics[width=16cm]{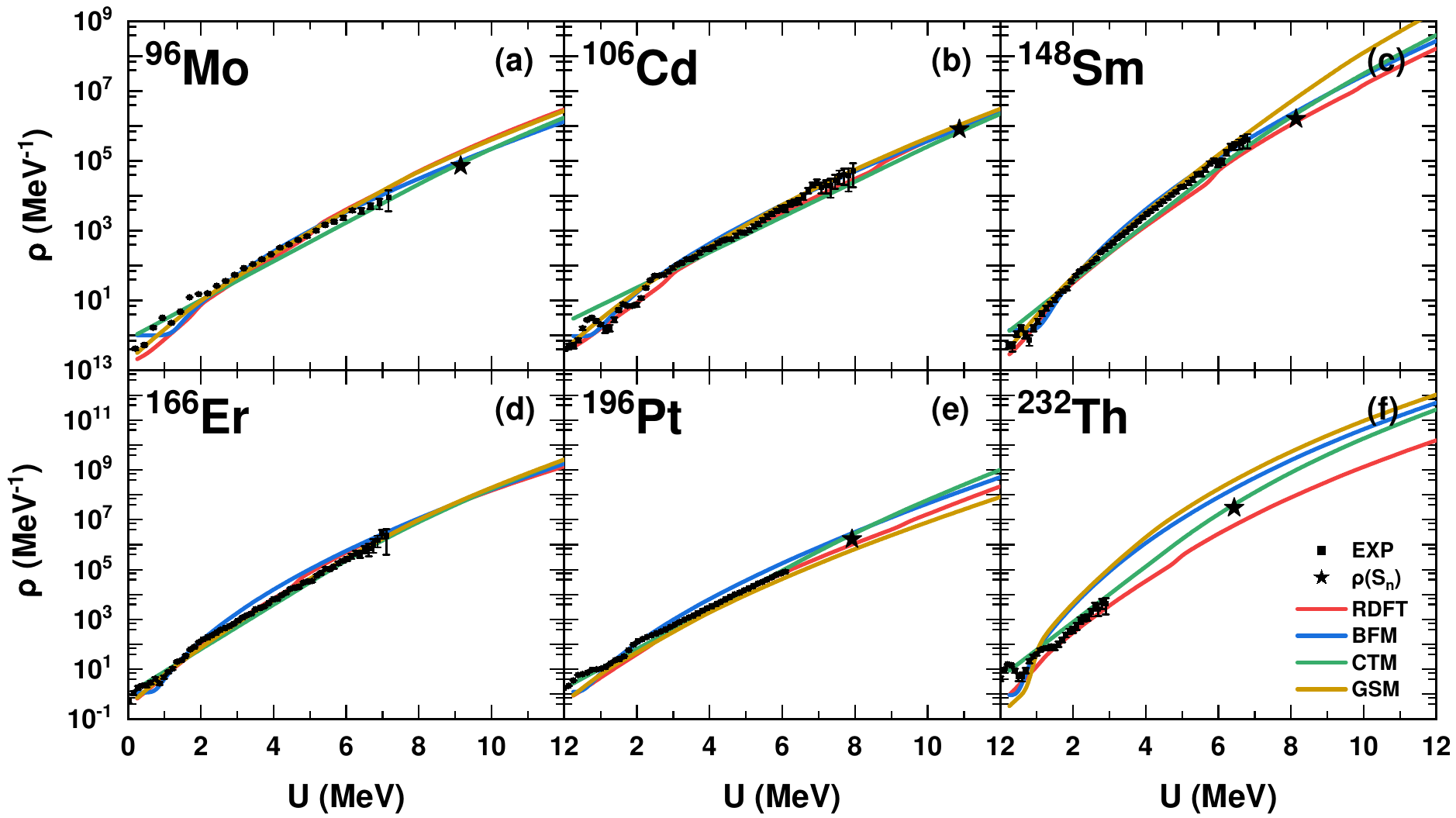}
\caption{\justifying Comparison of NLDs calculated using the RDFT-based statistical method with results from different phenomenological models in TALYS code \cite{EPJA2023596} and experimental data. The asterisk indicates the experimental level density at the neutron separation energy.}
\label{fig7}
\end{figure*}

The rotational enhancement factor shown in Fig. \ref{fig:5(c)} is determined, as in Ref. \cite{EPJA55249}, by the ratio of the state densities calculated with and without rotational effects:
\begin{equation}
K_{\text{rot}} = \frac{\sum_{J} (2J + 1)\rho(U, J)}{\omega(U)}, \label{eq:23}
\end{equation}
where $\omega(U)$ is given by Eq. (\ref{eq:3}), and $\rho(U, J)$ is given by either Eq. (\ref{eq:11}) or Eq. (\ref{eq:12}). The factor $2J+1$ accounts for the degeneracy of states with respect to the projection $M$ of the total angular momentum onto the laboratory $z$-axis. In the low excitation energy region, nuclear rotation is strongly suppressed due to pairing correlations. Meanwhile, both the magnitude and evolution of the rotational enhancement factor closely resemble the behavior of the perpendicular spin cut-off parameter shown in Fig. \ref{fig:5(b)}. This observation is consistent with the theoretical analysis presented in Ref. \cite{EPJA55249}. Figure \ref{fig:5(d)} shows that the vibrational enhancement factor differs among the various deformations. This difference originates from temperature variations. At the same excitation energy, higher temperatures lead to stronger vibrational enhancement.

\subsection{Nuclear level density}

The nuclear level density calculated within the RDFT-based statistical approach depends on nuclear structure information, such as single-particle energy levels, pairing gaps, and nuclear deformation. To investigate the influence of different relativistic effective interactions on the level density, Fig. \ref{fig6} presents the level densities obtained using the two commonly adopted RDFT effective interactions DD-ME2 and DD-PC1 \cite{PRC71024312}, together with available experimental data extracted using the Oslo method \cite{PRC73034311,PRC87014319,PRC65044318,PRC63044309,PRC90054330,PRC88024307} for comparison.

Overall, the level densities obtained with the two effective interactions exhibit very similar evolution trends as a function of excitation energy, and both reproduce the known experimental level densities reasonably well in the low excitation energy region. It should be noted that compared with DD-ME2, DD-PC1 generally produces stronger pairing gaps. Consequently, pairing correlations are broken at higher excitation energies in the DD-PC1 calculations.

Figure \ref{fig6} also presents a comparison between the level densities calculated using the RDFT-based statistical method and those obtained from a statistical model based on MSk7 as well as a combinatorial model based on BSk14. For $^{96}\mathrm{Mo}$, $^{148}\mathrm{Sm}$ and $^{196}\mathrm{Pt}$, the BSk14 combinatorial model generally predicts higher level densities than the RDFT results. For $^{166}\mathrm{Er}$ and $^{232}\mathrm{Th}$, the BSk14 results are higher than those of RDFT at excitation energies below about 5 MeV, and then gradually approach the RDFT predictions at higher excitation energies. The statistical results based on the MSk7 interaction are generally close to the RDFT results. However, with the exception of $^{148}\mathrm{Sm}$, the remaining nuclei exhibit smaller critical excitation energies than those obtained with RDFT. Overall, the statistical method based on RDFT performs comparably to the other two microscopic models in describing nuclear level densities.

Figure \ref{fig7} presents a comparison between the level densities calculated using the DD-ME2 effective interaction and those obtained from various phenomenological models. It can be seen that all models exhibit similar overall increasing trends of the level density as a function of excitation energy. Owing to the parameterization based on experimental data, phenomenological models are generally able to reproduce the experimental level density at the neutron separation energy. It is noteworthy that the RDFT-based microscopic statistical approach is also capable of reasonably reproducing the experimental level density at the neutron separation energy. In the excitation energy region below the neutron separation energy, the agreement between different models and experimental data is comparable, and the discrepancies among the models are relatively small. This indicates that the RDFT-based microscopic statistical approach has a similar overall ability to describe the experimental data as phenomenological models.

\begin{figure}
\centering
\includegraphics[width=\columnwidth]{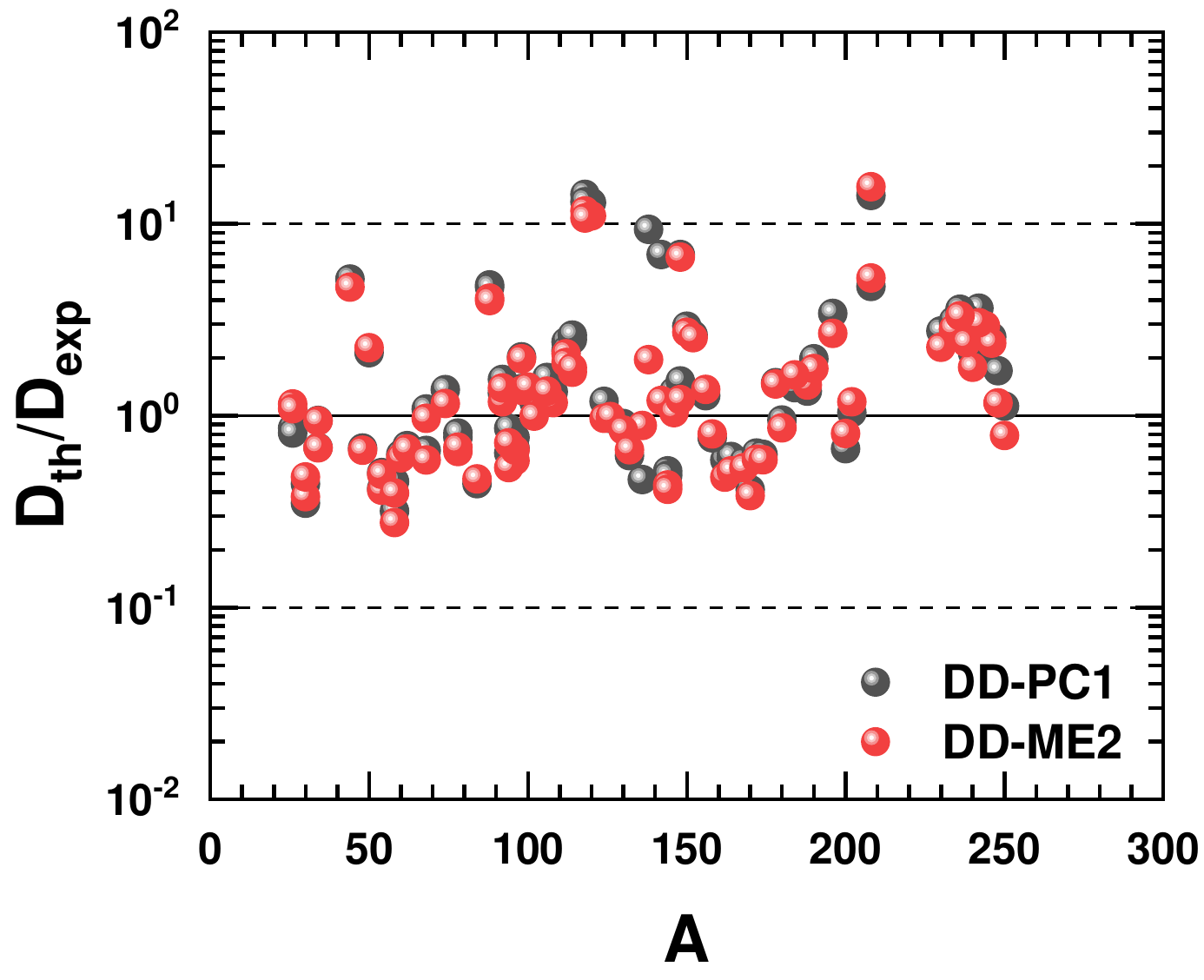}
\caption{\justifying Ratios of the $s$-wave and $p$-wave neutron resonance spacings calculated using the RDFT-based statistical method to the corresponding experimental data.}
\label{fig8}
\end{figure}

\subsection{Neutron resonance spacing}
To further assess the reliability of the RDFT-based statistical approach at the neutron separation energy, the $s$-wave and $p$-wave neutron resonance spacings were systematically analyzed. In this analysis, it was assumed that positive and negative parities are equally distributed. The agreement between theoretical and experimental results was quantitatively evaluated using the logarithmic root mean square deviation factor $f_{\text{rms}}$ defined as:
\begin{equation}
f_{\text{rms}} = \exp\left[ \frac{1}{N} \sum_{i=1}^{N} \left( \ln \frac{D^{i}_{\mathrm{th}}}{D^{i}_{\mathrm{exp}}} \right)^2 \right]^{1/2}, \label{eq:23}
\end{equation}
where $D^{i}_{\mathrm{th}}$ and $D^{i}_{\mathrm{exp}}$ denote the theoretical and experimental neutron resonance spacings, and $N$ is the number of nuclei considered. The distributions of the ratios ${D_{\mathrm{th}}}/{D_{\mathrm{exp}}}$ are shown in Fig. \ref{fig8}.

For the 67 even–even nuclei with available experimental s‑wave neutron resonance spacing data \cite{NDS20093107}, the present RDFT-based calculations yield $f_{\text{rms}} = 2.29$ with DD-ME2 and $f_{\text{rms}} = 2.56$ with DD-PC1. This accuracy is significantly better than that obtained with the relativistic combinatorial approach without pairing correlations, which yields $f_{\text{rms}} = 3.62$ for 66 nuclei \cite{NST34141}, and is comparable to the non-relativistic microscopic statistical method, for which $f_{\text{rms}} = 2.14$ is obtained for 278 nuclei \cite{NPA69595}. However, it remains larger than the values achieved by the phenomenological BFM, with $f_{\text{rms}} = 1.78$ for 295 nuclei \cite{NPA6381}, and by the QRPA-based boson expansion method, which gives $f_{\text{rms}} = 1.65$ for 48 nuclei \cite{PLB137989}. Among the 67 nuclei considered, 18 nuclei have available experimental data for $p$-wave neutron resonance spacings. For these nuclei, $f_{\text{rms}} = 2.84$ with DD-ME2 and $f_{\text{rms}} = 2.93$ with DD-PC1. When the $s$- and $p$-wave neutron resonance spacing data of the 67 nuclei are considered simultaneously, $f_{\text{rms}} = 2.41$ for DD-ME2 and $f_{\text{rms}} = 2.64$ for DD-PC1. These values are comparable to those obtained with the non-relativistic combinatorial approach including pairing correlations, which yields $f_{\text{rms}} = 2.3$ for 289 nuclei \cite{PRC064307}.

In summary, the RDFT-based microscopic statistical approach is capable of reproducing the experimental neutron resonance spacings with an accuracy comparable to that of existing non-relativistic microscopic models. Among the two effective interactions considered, DD-ME2 performs slightly better than DD-PC1 in a statistical sense, although the difference between them is not significant.

\section{\label{sec:4}SUMMARY AND PROSPECTS}

In this work, the nuclear level density has been systematically investigated within a microscopic statistical approach based on RDFT. The method is built upon self-consistent single-particle levels obtained from RDFT calculations and provides a unified statistical description of the dependence of the nuclear level density on excitation energy and spin.

In this approach, the spin cut-off parameter is calculated microscopically based on the single-particle levels. This method effectively corrects the systematic overestimation commonly encountered in the semi-empirical formulas at low excitation energies, and also overcomes the unreliability of the global empirical formula when extrapolated to high excitation energies.

Using the shape-coexisting nucleus $^{98}\mathrm{Sr}$ as an example, the dependence of nuclear state density and associated collective enhancements on nuclear deformation has been investigated. The results indicate that differences in the single-particle levels and pairing gaps at different deformations are the primary microscopic origin of variations in nuclear state density and collective enhancements. In particular, rotational enhancement is also related to nuclear deformation, and the degree of rotational enhancement varies with different deformations.

Furthermore, systematic calculations for even–even nuclei have been performed using the RDFT-based statistical method. The calculated results are compared with those from various phenomenological and microscopic models, as well as with available experimental data. These comparisons indicate that the present approach can reasonably reproduce experimental $s$- and $p$-wave neutron resonance spacings and nuclear level densities. Although some deviations exist among the different NLD models, their overall trends remain consistent. This consistency further confirms the reliability of the RDFT-based statistical approach in studies of nuclear level density.

It should be noted that the temperature dependence of the single-particle levels has not been explicitly considered in the present calculations. In the future, incorporating temperature-dependent density functionals into the microscopic statistical framework is expected to improve the accuracy of nuclear level density predictions and extend their applicability to higher excitation energies. Furthermore, machine learning offers a promising approach, having already been applied to phenomenological nuclear level density models \cite{ARI2021169109583,PRC20241094044325}. Future efforts could combine machine learning with the microscopic statistical model, building upon recent studies of nuclear charge densities \cite{NST20223312153,PRC20241101014308,NST202637593}.

\begin{acknowledgments}

The work of Z.C.W., P.X.D. and J.L. was supported by the National Natural Science Foundation of China (Grants No.12475119 and No.12447101) and Key Laboratory of Nuclear Data Foundation (JCKY2025201C154). J.L. also thanks Peng Huanwu Visiting Professor Program at the Institute of Theoretical Physics of Chinese Academy of Sciences. The work of S.G.Z. was supported by the National Key R\&D Program of China (Grants No. 2024YFE0109800 and No. 2023YFA1606500), the National Natural Science Foundation of China (Grants No. 12447101, No. 12375118, No. 12435008, and No. W2412043), and the Strategic Priority Research Program of Chinese Academy of Sciences (Grants No. XDB0920000 and No. XDB1550000).
\end{acknowledgments}

%\nocite{*}

\bibliography{apssamp}% Produces the bibliography via BibTeX.

\end{document}